\documentclass[prl, twocolumn, superscriptaddress]{revtex4-1}

\usepackage{amssymb,amsmath,xcolor,graphicx,xspace,colortbl,textcomp, rotating} 
\usepackage{amsmath}  
\usepackage{amssymb}  
\usepackage{amssymb,amsmath,xcolor,graphicx,xspace,colortbl,textcomp, rotating}  
\usepackage{bm}  
\usepackage{color}  
\usepackage{dcolumn}  
\usepackage{graphicx}  
\usepackage[cp1251]{inputenc}  
\usepackage{multirow}  
\usepackage{soul}  
\usepackage[colorlinks]{hyperref}  
\graphicspath{{curie170504AVAK_graphics/}{curie170504AVAK_tcache/}{curie170504AVAK_gcache/}}
\DeclareGraphicsExtensions{.pdf,.eps,.ps,.png,.jpg,.jpeg}
\graphicspath{{curie170418_KVAK_graphics/}{curie170418_KVAK_tcache/}{curie170418_KVAK_gcache/}}
\DeclareGraphicsExtensions{.pdf,.eps,.ps,.png,.jpg,.jpeg}

\definecolor{orange}{rgb}{1,0.7,0.4}
\definecolor{green}{rgb}{0.3,1,0.3}

\begin{document}
\title{The Ferromagnetism in the Vicinity of Lifshitz Topological Transitions}
\author{P.S.~Grigoryev}
\affiliation {Spin Optics Laboratory, St. Petersburg State University, Ulyanovskaya 1, 198504 St. Petersburg, Russia}
\email{f.grigoriev@spbu.ru}
\author{M.M.~Glazov}
\affiliation {Ioffe Institute, 26 Polytechnicheskaya str., 194021, St.-Petersburg, Russia}
\author{A.V.~Kavokin}
\affiliation {CNR-SPIN, Tor Vergata, Viale del Politecnico 1, I-00133 Rome, Italy}
\affiliation {School of Physics and Astronomy, University of Southampton, Southampton SO17 1BJ, United Kingdom}
\affiliation {Spin Optics Laboratory, St. Petersburg State University, Ulyanovskaya 1, 198504 St. Petersburg, Russia}
\author{A.A.~Varlamov}
\affiliation {CNR-SPIN, Tor Vergata, Viale del Politecnico 1, I-00133 Rome, Italy}

\date{
\today 
}
\begin{abstract}We show that the critical temperature of a ferromagnetic phase transition in a quasi-two-dimensional hole gas confined in a diluted
magnetic semiconductor quantum well strongly depends on the hole chemical potential and hole density. The significant variations of the the Curie temperature occur close to the Lifshitz topological transition points where the hole Fermi surface acquires additional  components of topological connectivity due to the filling of excited size-quantization subbands. The model calculations demonstrate that
the Curie temperature can be doubled by a small variation of the gate voltage for the CdMnTe/CdMgTe quantum well based device.
\end{abstract}

\pacs{75.75.-c, 75.50.Pp}

\maketitle


Quantum wells based on diluted magnetic semiconductors have been attracting attention for several decades due to their unusual magneto-optical properties
~\cite{Furdyna-JAP1988, RevModPhys.78.809, RevModPhys.86.187}. One of the most interesting observed effects that is potentially promising
for the applications in semiconductor spintronics is the carrier-induced ferromagnetism~\cite{dietl:R3347, PhysRevLett.79.511}.
The magnetic ordering of the spins of magnetic ions may happen due to their exchange interaction with spins of delocalised carriers, typically holes. In
particular, the doping of III-V semiconductor\ by Mn$^{2 +}$ ions results in formation of the acceptor states~\cite{Kitchen-Nature2006},
or even impurity bands in the case of heavy doping~\cite{Ohya-NaturePhys2011, Dobrowolska-NatureMat2012, Gray-NatureMat2013, Ishii-PRB2016}. The holes coming
from these acceptors help achieving the ferromagnetic ordering at elevated temperatures~\cite{Jungwirth-PRB2005}.

Ferromagnetic ordering of spins of magnetic ions at relatively low concentrations (several
percent) can be described in terms of the  (RKKY) mechanism. The record Curie temperature achieved due to this mechanism in III-V semiconductors doped with
Mn is about $150$\,K~\cite{Jungwirth-PRB2005} for the best quality samples. The most optimistic
theoretical estimates predict the ferromagnetism in GaMnAs at moderate Mn concentrations even at room-temperature~\cite{Jungwirth-PRB2005}.
Yet, it turns out that the significant amount of holes is compensated due to interstitial Mn incorporation~\cite{Agrinskaya-SSC2014}
that reduces the ferromagnetic transition temperature. 

In contrast to III-V semiconductors, in II-VI semiconductors Mn ions do not form acceptor states. Therefore they
weakly affect the band structure of the host semiconductor. This is why, describing the coupling of magnetic ion spins with carriers in II-VI semiconductors, one can rely on the symmetry based band description. This allows for a straightforward description of the electronic properties of diluted magnetic quantum wells. 

In this Letter, we show that the critical conditions for the ferromagnetic phase transition in II-VI semiconductor quantum wells containing a low concentration
of magnetic ions become extremely sensitive to the chemical potential
of the hole gas in the vicinity of topological transition points in the valence band. This opens way for an efficient control of the Curie temperature in such structures by an external bias. Switching on and off the ferromagnetism may be achieved by a weak variation of voltage applied to a properly designed diluted magnetic quantum well. This property, based on the fundamental physics of Lifshitz topological transitions, may open way to realisation of a new class of spintronic devices.

It is well known that once the electronic chemical potential crosses the bottom of one of the size-quantization subbands in a quantum
well the Fermi surface acquires a new component of topological connectivity. This transition is a particular case of the Lifshitz topological transition~\cite{Blanter1994}.
It has been recently demonstrated that such transformations are accompanied by the spikes in the entropy per particle as well as by the spikes in the temperature
derivative of the chemical potential of the electron or hole gas~\cite{Varlamov-PRB2016}. Below we show that the magnetic susceptibility of the system experiences similar spikes in the vicinity of the Lifshitz transition points. This
leads to a very strong variation of the Curie temperature with a small variation of the chemical potential of the hole gas that may be achieved by the application
of bias. 

\begin{figure}[htbp]
\centering
\includegraphics [width=7cm]{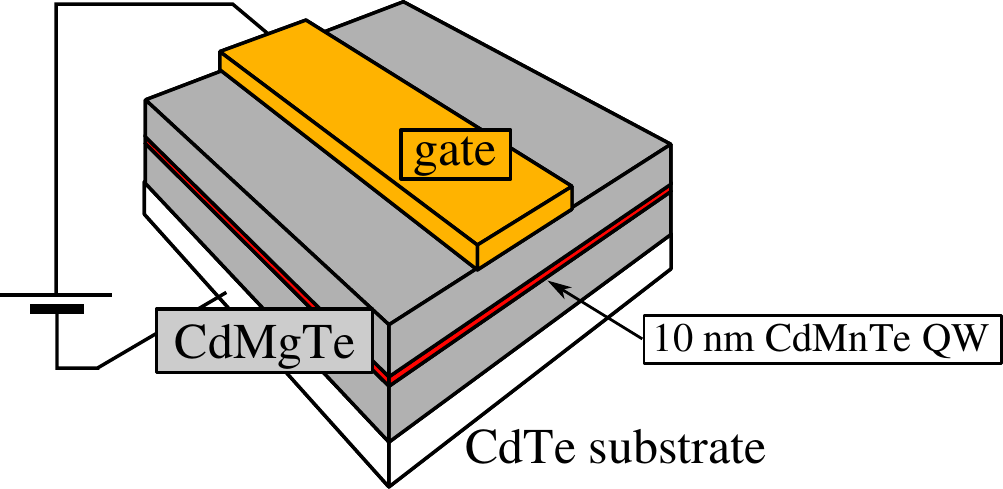}
\vspace{-0.3cm}
\caption{%
(color online) The considered model structure contains a single CdMnTe quantum well sandwiched between CdMgTe barriers. The gate voltage applied to the structure controls the concentration of holes in the quantum well.}
\vspace{-0.6cm}
\label{fig2}\end{figure}

Let us consider a CdMnTe/CdMgTe quantum well, embedded in a gated structures schematically shown in Figure~\ref{fig2}.
The gate voltage $U_{g}$ applied to the structure controls the density of the two-dimensional hole gas (2DHG) confined in the QW, $n_{h}$. Within the linear approximation, $n_{h} =C U_{g}/\vert e\vert $, where $e$ is the electron charge and $C$ is the structure capacitance. In what follows, for simplicity we assume that the capacitance $C$ is independent of the gate voltage and it is governed by the geometry of the structure rather than by the density of states~\cite{daviesBOOK,doi:10.1063/1.99649}.

A ferromagnetic ordering of the Mn-ions magnetic momenta is possible due to the RKKY interaction between the ion spins
mediated by the 2DHG~\cite{Furdyna-JAP1988,PhysRevLett.79.511,dietl:R3347,RevModPhys.78.809,RevModPhys.86.187}.
Due to the exchange interaction between the hole and Mn spins the fluctuations of the latter give rise to the hole spin polarization which, in turn, provides
the positive feedback to the Mn spin system. Quantitatively, in bulk semiconductors the exchange interaction of $d$-electrons of Mn ions with the hole spin is described by the effective Hamiltonian~\cite{merkulov99,RevModPhys.78.809,RevModPhys.86.187}
\begin{equation}
\label{exch}
\mathcal H_{\rm exch} = -\sum_i J_{pd}\delta(\bm r - \bm R_i)(\bm J\cdot \bm I_i),
\end{equation}
where $J_{p d}$ is a constant, $\bm J$ are the matrices of the hole angular momentum $3/2$, $\bm I_i$ are the Mn spin operator described by the spin-$5/2$ matrices, the summation is carried out over all Mn ions enumerated by the subscript $i$. In Eq.~\eqref {exch} $\bm r$ 
and $\bm R_i$ are the position vectors of the hole
and $i$th Mn, respectively. In a quantum well structure Eq.~\eqref {exch} should be averaged over appropriate size quantization states. Hereafter we consider only heavy- and light-hole states and disregard the spin-orbit split-off branch due to significant energy separation between the $\Gamma_8$ and $\Gamma_7$ valence bands~\cite{Lawaetz-PRB1971}. We assume that the system shown in Fig.~\ref{fig2}
contains a symmetric quantum well grown along $z \parallel [001]$ axis. Diagonalizing the Hamiltonian~\eqref{exch} we obtain the series of the heavy- and light- hole subbands, $h h \nu $ and $l h \nu $, respectively, with $\nu  =1 ,2 ,\ldots $ with corresponding envelope functions $\varphi _{h h \nu } (z)$, $\varphi _{l h \nu } (z)$. The exchange interaction of the heavy- and light holes with magnetic ions is described by
\begin{subequations}
\label{exch:QW}
\begin{equation}
\mathcal H_{\rm hh} = -\sum_i J_{hh\nu,i} ( s_{hh,z}  I_{i,z}) \delta(\bm \rho - \bm \varrho_i),\label{exch:hh}
\end{equation}
\begin{multline}
\mathcal H_{\rm lh} = -\sum_i J_{lh\nu,i} [s_{lh,z} I_{i,z} + 2(s_{lh,x}I_{i,x} +s_{lh,y}I_{i,y})] \\ \times \delta(\bm \rho - \bm \varrho_i).
\end{multline}
\end{subequations}
Here $J_{h h \nu  ,i} =3 J_{p d} \vert \varphi _{h h \nu } (z_{i})\vert ^{2}$, $J_{l h \nu  ,i} =J_{p d} \vert \varphi _{l h \nu } (z_{i})\vert ^{2}$, $\bm \rho$ and $\bm \varrho_i$ are the in-plane
position vectors of the hole and $i$th Mn ion, $s_{h h ,z}$ is the $z$-component of the heavy-hole pseudospin ($s_{h h ,z} = \pm 1/2$ for $J_{z} = \pm 3/2$), and $s_{l h}$ is the light-hole pseudospin. Note that in quantum well structures the
exchange interaction is anisotropic both for the heavy and light holes. A particularly strong anisotropy is found for the heavy holes where the interaction described by Eq.~\eqref{exch:hh} acquires the Ising form.

In order to illustrate the appearance of the magnetic phase transition and estimate the Curie temperature $T_{c}$ in the Mn spin system we use the mean field approach. We take into account only interaction between $z$-spin components of Mn ions and holes, and represent the thermodynamic potential in the form
\begin{eqnarray}
\Phi  &  = & \frac{I_{z}^{2}}{2 \chi ^{(M n)}} +\frac{s_{h h ,z}^{2}}{2 \chi ^{(h h)}} +\frac{s_{l h ,z}^{2}}{2 \chi ^{(l h)}} -  I_{z} (J_{hh} s_{h h ,z} -J_{lh} s_{l h ,z}) \nonumber \\
 &  &  +\mu _{B} B_{z} (g_{M n}  I_{z} +g_{h h}  s_{h h ,z} +g_{l h}  s_{l h ,z})\;\text{,}\; \label{Phi:Land}
\end{eqnarray}
where $g_{M n}$, $g_{h h}$ and $g_{l h}$ are the Mn and hole $g$-factors, respectively, $\mu _{B}$ is the Bohr magneton, $B_{z}$ is $z$-component of the external magnetic field, $I_{z}$, $s_{h h ,z}$ and $s_{l h ,z}$ are the spin densities of Mn, heavy- and light- holes, $J_{h h}$ and $J_{l h}$ are the averaged exchange interaction constants for the corresponding hole states~\cite{note:J}. Here $\chi ^{(M n)}$ is the non-interacting susceptibility of Mn ions,
\begin{equation}\chi ^{(M n)} =\frac{I (I +1) n_{M n}}{3 k_{B} T}\;\text{,}\; \label{chi:Mn:0}
\end{equation}  with $I =5/2$ being the Mn spin and $n_{M n}$ is the density of Mn ions on the sample, $\chi ^{(h h)}$ and $\chi ^{(l h)}$ are the static susceptibilities of the heavy and light holes. Note that the susceptibilities defined here provide the link between
the spin density and the field-induced level splitting, e.g., $I_{z} = -\chi ^{(M n)} g_{M n} \mu _{B} B_{z}$. Minimizing $\Phi $ with respect to $I_{z}$, $s_{h h ,z}$, and $s_{l h ,z}$ allows us to obtain the effective susceptibility of the Mn spin system as:
\begin{equation}\tilde{\chi }^{(M n)} =\frac{\chi ^{(M n)}}{1 -(J_{h h}^{2} \chi ^{(h h)} +J_{l h}^{2} \chi ^{(l h)}) \chi ^{(M n)}}\;\text{.}\; \label{chi:Mn:1}
\end{equation}  Since, in accordance
to the fluctuation-dissipation theorem, the spin susceptibility is proportional to the spin-spin correlation function~\cite{ll5_eng},
$\tilde{\chi }^{(M n)} =\mathcal{S}  \langle I_{z}^{2} \rangle /k_{B} T$, where $\mathcal{S}$ is the normalization area, the divergence of the susceptibility corresponds to the phase transition point.
It follows from Eqs.~\eqref {chi:Mn:0}-\eqref {chi:Mn:1} that the divergence occurs where the denominator vanishes, which yields the self-consistent equation for the Curie temperature~\cite{dietl:R3347,RevModPhys.78.809,RevModPhys.86.187}
\begin{equation}
T_{c} =\frac{I (I +1) n_{\text{Mn}}}{3 k_{B}} \left (J_{h h}^{2} \chi ^{(h h)} +J_{l h}^{2} \chi ^{(l h)}\right ) . \label{Crit_Temp}
\end{equation}  

We follow the standard approach \cite{ll5_eng} to analyze the applicability of the mean-field treatment. We consider a fluctuation of the thermodynamic potential given by:
\begin{equation}
\label{Phi:Land:q}
\delta \Phi = g\left(\frac{\partial s_{z}}{\partial \bm r}\right)^2+ \frac{b I_z^4}{2\chi^{(\rm Mn)}},
\end{equation}
where $g$ and $b$ are the constants. The parameter $b$ can be found from development of the Brillouin function in a series and retaining cubic terms, $b =(I^{2} +I +1/2)/[10 I(I +1) n_{M n}^{2}]$. For the long wavelength fluctuations with the wavevector $\bm q$ the effective Mn susceptibility reads as [cf. Eq.~ 
\eqref {chi:Mn:1}]:
\begin{equation}
\label{chi:Mn:1:q}
\tilde \chi^{(\rm Mn)}_{\bm q} \approx \frac{\chi^{(\rm Mn)}}{1+(ql_s)^2-T_c/T},
\end{equation}
Here $l_{s} =\sqrt{g \chi ^{(h)}}$. The approximate equality holds for $T -T_{c} \ll T_{c}$ and $q l_{s} \ll 1$. Equation~\eqref{chi:Mn:1:q}
 allows one to obtain
the correlation radius of fluctuations (above $T_{c}$) in the standard form
\begin{equation}r_{c} =\frac{l_{s}}{\sqrt{1 -T_{c}/T}}\;\text{.}\; \label{rc}
\end{equation}
Note, that the applicability of the mean field approximation is restricted by the requirement of the relative weakness of the spin density fluctuations. Namely, the mean square fluctuation of $I_{z}$ per unit square, $\delta  I_{z}^{2} \sim k_{B} T_{c} \tilde{\chi }^{(M n)}/r_{c}^{2}$, must be small with respect to the average value of the spin density square $\;I_{z}^{2}\text{.}$The latter is obtained in the mean field approximation, i.e. by means of minimization of the functional $\Phi  +\delta  \Phi $ (see Eqs.~\eqref {Phi:Land} and \eqref {Phi:Land:q}) omitting the fourth-order term: $I_{z}^{2} =(T_{c} -T)/(2 b T_{c})\text{.}$ This analysis brings us to the conclusion, that the mean field approximation is valid in the temperature range given by
\begin{equation}\mathrm{G} \mathrm{i} \ll \frac{\vert T -T_{c}\vert }{T_{c}} \ll 1\;\text{,}\; \label{crit}
\end{equation}
where the quantity
$\mathrm{G} \mathrm{i} =\left (n_{M n} l_{s}^{2}\right )^{ -1}$ plays the role of Ginzburg-Levanyuk number~\cite{ll5_eng} in the two-dimensional case under consideration. One can see that the inequality \eqref {crit} is fulfilled in diluted magnetic CdTe-based quantum wells. Indeed, the
parameter $l_{s}$ for free holes at low temperatures  can be estimated as $l_{s} \sim k_{F}^{ -1}$, where $k_{F}$ is the hole wavevector. Hence, the condition $\mathrm{Gi} \ll \mathrm{1}$ \ means that the density of holes is small in comparison to the density of Mn ions, that
is always the case. For example, in our case minimal 1\% Mn concentration corresponds to $\approx10^{14}$\,cm$^{-2}$ and hole concentration usually does not exceed $10^{13}$\,cm$^{-2}$.

For the non-interacting hole gas the susceptibility can be written as~\cite{ll5_eng}:
\begin{equation}
\chi ^{(h)} = -\frac{1}{4 \mathcal{S}} \frac{ \partial ^{2}\Omega ^{(j)}}{ \partial \mu ^{2}} =\frac{1}{4} \frac{ \partial n_{j}}{ \partial \mu }\;\text{.}\; \label{chi:non:int}
\end{equation}  
where $\Omega ^{(j)} \equiv \Omega ^{(j)} (\mu )$ is the (Grand) thermodynamical potential, $\mu $ is the chemical potential, $n_{j}$ is the density of the corresponding hole states, $j =h h$ or $l h$. The latter can be conveniently expressed through the density of the hole states, $g_{h} (E)$ as
\begin{equation}
n_{h} =\int _{0}^{\infty }\frac{g_{h} (E) d E}{\exp{\left(\frac{E -\mu }{k_{B} T}\right)} +1}, 
\label{integral}
\end{equation}
where the energy is reckoned from the $h h 1$ subband size quantization energy. In what follows Eqs.~\eqref {chi:non:int} and \eqref{integral} are used to numerically calculate the susceptibilities of the hole gas as functions of temperature and the hole density and to solve
Eq.~\eqref {Crit_Temp} in order to find the Curie temperature $T_{c}$ as a function of the hole density or gate voltage. 

Before presenting the numerical results, let us focus on the
simplified analytical model which takes into account only one type of holes but provides a clear physical picture of the effect of topological Lifshitz transitions
on the hole spin susceptibility and the ferromagnetic order of Mn spins. Let us represent the density of heavy-hole states as $g (E) =(m^{ \ast }/\pi  \hbar ^{2}) \sum _{\nu }\Theta  (E -E_{h h \nu })$, where $m^{ \ast }$ is the heavy-hole effective mass, $E_{h h \nu }$ are the energies of the size-quantized subbands, and $\Theta  (E)$ is the Heaviside step function~\cite{daviesBOOK,Varlamov-PRB2016}. Furthermore, let us assume that relevant temperatures are low enough so that the thermal broadening
in Eq.~\eqref {integral} can be disregarded and the heavy-hole susceptibility can be recast in the form
\begin{equation}\chi ^{(h h)} =\frac{g (\mu )}{4} =\frac{m^{ \ast }}{4 \pi  \hbar ^{2}} \sum _{\nu }\Theta  (\mu  -E_{h h \nu })\;\text{.}\; \label{chi:non:int:2}
\end{equation}  It follows
from Eq.~\eqref {Crit_Temp} that the Curie temperature $T_{c} \propto J_{h h}^{2} n_{M n} \chi ^{(h h)} (\mu )$ as a function of the hole chemical potential $\mu $ demonstrates a steplike increase as soon as $\mu $ touches the consequent heavy-hole subband. This is because the hole Fermi surface acquires a new component of topological
connectivity giving rise to the Lifshitz phase transition. It is accompanied by the steplike increase of the density of states since with further increase of the chemical potential more subbands start to get filled~\cite{Blanter1994}.
Hence, the Curie temperature increases by a certain value at the point of the Lifshitz transition. These results are corroborated by the numerical analysis
below. We note, however, that Eq.~\eqref {chi:non:int:2} cannot be used in the narrow vicinity of the topological transition point where $\vert \mu  -E_{h h \nu }\vert  \lesssim k_{B} T$. The kinks in the $T_{c} (\mu )$ dependence predicted by Eqs.~  \eqref {Crit_Temp} and \eqref {chi:non:int:2} are smoothed-out, as shown below, due to
thermal spread of the electron distribution function. 



\begin{figure}[htbp]
\centering
\vspace{-0.4cm}
\includegraphics [scale=1]{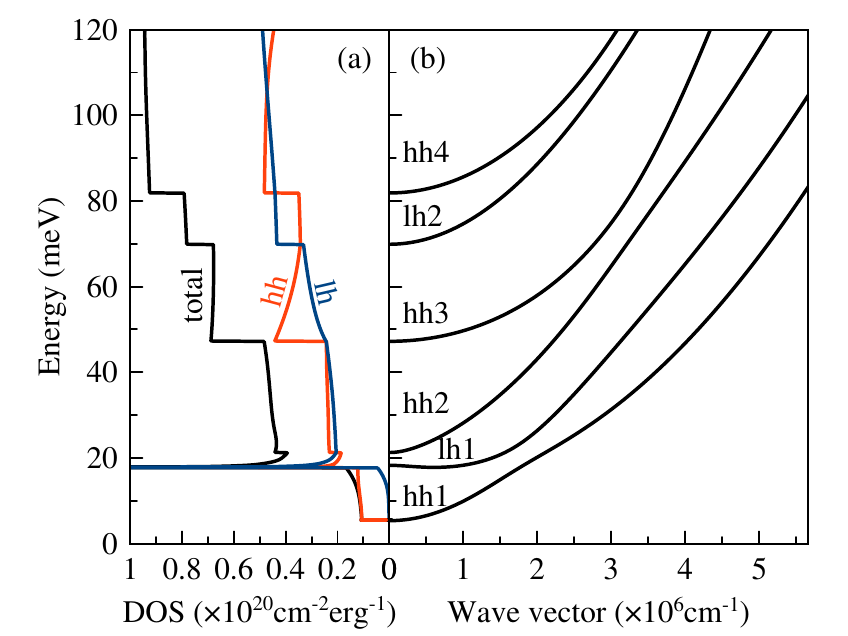}
\vspace{-0.8cm}
\caption{
(color online) (a) Density of hole states in 10-nm CdMnTe/CdMgTe QW (black) and the partial contributions of the heavy- (red) and light-hole (blue) states to the density of states. (b) Sub-bands of size quantisation in 10-nm CdMnTe/CdMgTe QW. In calculation we used 5 basic functions for heavy and light holes, and spherical approximation for the Luttinger Hamiltonian with the parameters $\gamma _{1} =5.7$,$\gamma _{2} =\gamma _{3} =1.7$~\cite{Lawaetz-PRB1971} and the barrier height of $120$~meV.}
\vspace{-0.2cm}
\label{fig4}\end{figure}

Figure~\ref{fig4} shows the results of numerical calculations for the 10-nm thick CdMnTe/CdMgTe QW. In order to find the energy dispersion of the heavy- and light-holes ($hh\nu$ and $lh\nu$ subbands, respectively) in the QW and the corresponding effective exchange interaction constants $J_{hh\nu}$, $J_{lh\nu}$ we have numerically diagonalized the Luttinger Hamiltonian. We made use of the fact that at zero in-plane hole wavevector $\bm k_{\parallel}$ the states with $|J_z|=3/2$ (heavy-holes) and $|J_z|=1/2$ (light-holes) are decoupled. Hence, we, at first, found the heavy- and light-hole functions, $\varphi_{hh\nu}(z)$, $\varphi_{lh\nu}(z)$ at $\bm k_\parallel =0$. Second, we represented the hole wavefunctions at $\bm k_\parallel \ne 0$ as linear combinations of $\varphi_{hh\nu}(z)$, $\varphi_{lh\nu}(z)$ and diagonalized the obtained matrix Hamiltonian. The calculated density of states and energy dispersions for the hole subbands are shown in Fig.~\ref{fig4}, panels (a) and (b) respectively.

In the studied range of energies the heavy- and light-hole states are substantially mixed due to the off-diagonal elements of the Luttinger Hamiltonian. It results in the energy non-parabolicity and anticrossing behavior of the dispersion curves presented in Fig.~\ref{fig4}(b). To illustrate the mixing in more detail we presented in Fig.~\ref{fig4}(a) to the total density of states the partial contributions of the heavy- and light-holes calculated as
\begin{equation}
\label{partial:dos}
g_{hh(lh)}(E) = \frac{1}{4\pi} \sum_{\nu} \frac{d k_\parallel^2}{d E} \mathcal C_{hh(lh)}^{(\nu)}(k_\parallel),
\end{equation}
where the summation is carried out over all dispersion branches at a given energy, $\mathcal C_{hh(lh)}^{(\nu)}(k_\parallel)$ is the fraction of the corresponding heavy ($hh$) or light ($lh$) hole state in the subband state $\nu$, $\mathcal C_{hh}^{(\nu)}(k_\parallel)+\mathcal C_{lh}^{(\nu)}(k_\parallel)=1$. Note that in the spherical approximation employed here the dispersion is isotropic in the QW plane.
\begin{figure}[htbp]\centering 
\centering
\includegraphics [width=8.6cm]{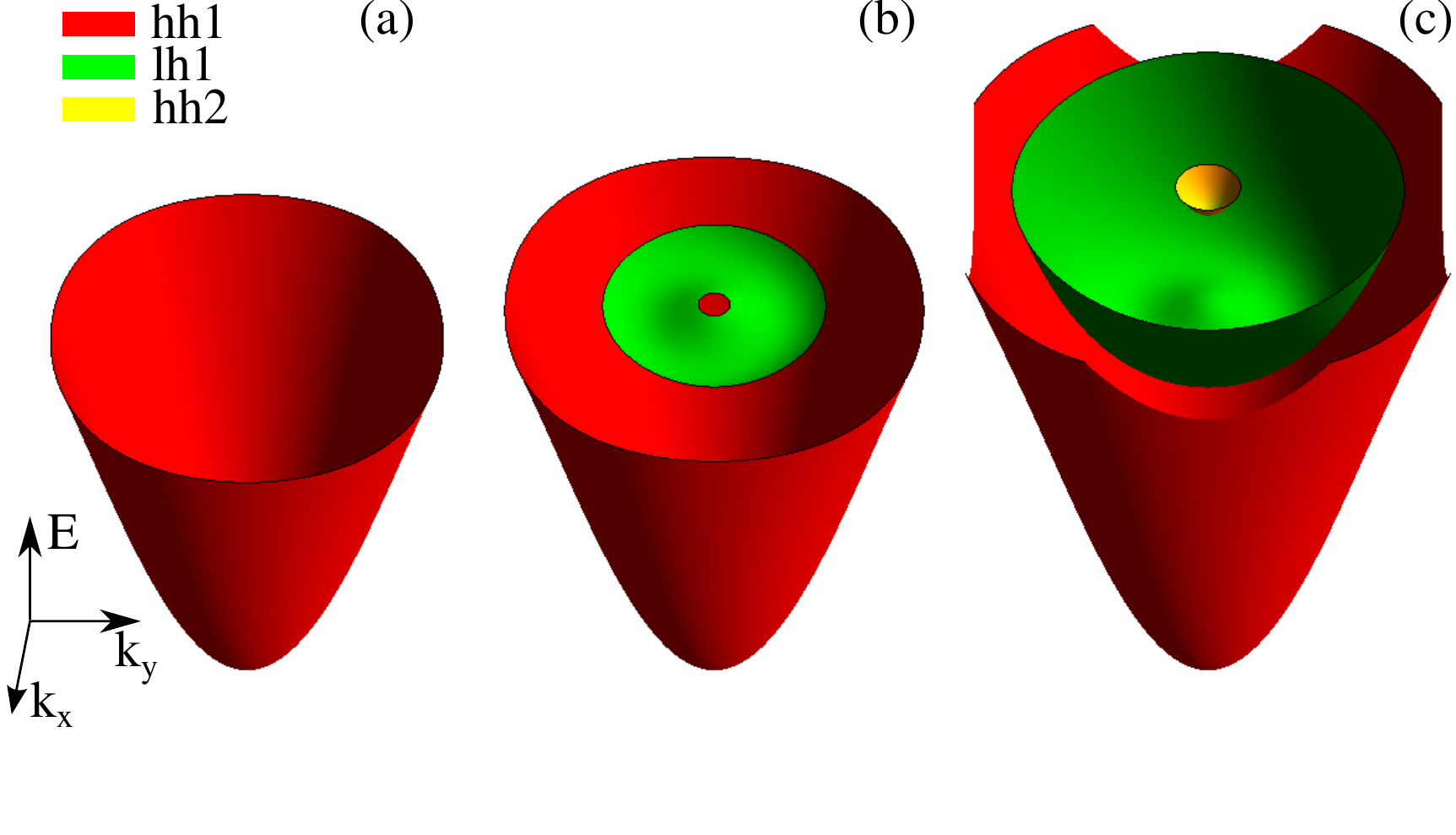}
\vspace{-1.2cm}
\caption{(color online) The schematic illustration of the hole dispersion surfaces cut by the Fermi energy in a with the Fermi energy below the second hole sub-band bottom (a), below the third hole sub-band bottom (b), and above the third hole sub-band bottom (c). }
\vspace{-0.6cm}
\label{fig5}\end{figure}  

The results of this calculation presented in Fig.~\ref{fig4}(a) show the step-like behavior of the density of states. Moreover, the total density of states as well as partial contributions of the heavy- and light-holes demonstrate sharp peaks followed by shallow minima in the vicinity of the onset of the first and second excited subbands, see the $lh1$ and $hh2$ subbands in Fig.~\ref{fig4}(b) and the density of states in the vicinity of $\approx 20$~meV hole energy in the panel (a). This result is a hallmark of the heavy-light hole mixing, which is the specific feature of the II-VI and III-VI semiconductors: For the parameters of the studied structure the ground light-hole subband and the excited heavy-hole subband are close in energy and are strongly mixed via the off-diagonal $\propto -\mathrm i k_\parallel \partial/\partial z$ terms of the Luttinger Hamiltonian. As a result of the subband repulsion the dispersion of the first excited subband is non-monotonic as seen in Fig.~\ref{fig4}(b), $E_{lh1} \approx \hbar^2 E_m+(k-k_0)^2/2m^*$,  where $E_m$ is the bottom of the first excited subband, $k_0$ is the wavevector corresponding to the subband bottom. As a result, the extremum loop is formed as it is seen in Fig.~\ref{fig5}, where in panels (a), (b) and (c) the cuts of energy dispersion surfaces at different energies are presented. The density of states has a one-dimensional-like singularity $\propto 1/\sqrt{E-E_m}$ in agreement with the numerical calculation. Note that the singularly can be strongly enhanced by the interface-induced heavy-light hole mixing resulting from the chemical bonds anisotropy in cubic crystalline lattices~\cite{aleiner}. At somewhat higher energies an effective gap is formed between the first and second excited subbands, which results in a minimum of the density of states.

\vspace{-0.4cm}
\begin{figure}[htbp]
\centering
\includegraphics [scale=1]{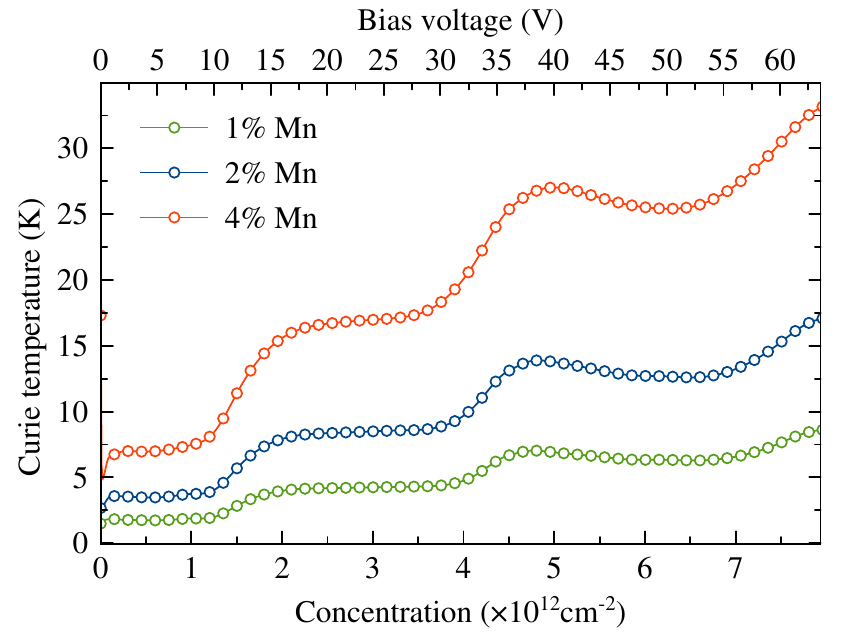}
\vspace{-0.8cm}
\caption{(color online) Dependence of the Curie temperature on the bias voltage and the hole concentration for three values of the Mn concentration in the QW, $J_{h h} =3 J_{l h} =3 \times 59$\,meV$\cdot$nm$^3$, device capacitance $C=20$\,nF/cm$^2$.}
\vspace{-1.3cm}
\label{fig6}\end{figure}  
\vspace{0.8cm}

Let us now discuss the Curie temperature dependence on the hole density/gate voltage in the QW structure presented in Fig.~\ref{fig6}. In the calculation we took the following values of the exchange interaction constants in Eq.~\eqref{chi:Mn:1} $J_{h h} =3 J_{l h} =3{J_{p-d}}$, with ${J_{p-d}}  =59$\,meV$\cdot$nm$^3$ being the $p-d$ exchange integral in the CdTe semiconductor~\cite{Furdyna-JAP1988,RevModPhys.78.809,RevModPhys.86.187}. In order to relate the gate voltage $V_g$ and the hole density $n$ we used the plane capacitor relation $n=CV_g/e$, where $e>0$ is elementary charge and the capacitance $C=20$\,nF/cm$^2$ ($C/e =1.25 \times 10^{11}$\,cm$^{ -2}$/V). Chosen capacitance is comparable to the real observable value in CdTe solar cells~\cite{Friesen-TSF2001}.

Figure~\ref{fig6} shows that each Lifshitz transition in the system is accompanied with a step-like increase in the Curie temperature. In particular, the Curie temperature doubles with the bias voltage changing between 10 and 15\,V. As expected, the Curie temperature is the higher the higher the Mn content in the QW is, which makes steps in the dependence of $T_c$ on the gate voltage/hole density more pronounced. Due to the thermal smoothing of the steps and peaks in the density of states, seen in Fig.~\ref{fig4}(a), the steps are smooth and, moreover, for the studied system the square-root singularity in the density of states in the vicinity of the first and second excited subbands is not reproduced in the $T_c$ dependence on the hole density. 

Interestingly, the Curie temperature slightly decreases with the increase in the hole density in the range of gate voltages from 40\,V to 50\,V. This corresponds to the hole chemical potential varying from 50\,meV to 70\,meV, approximately, where, in accordance with Fig.~\ref{fig4}(a) the heavy-hole fraction in the density of states decreases because the corresponding subband becomes more and more light-hole like. Since the exchange interaction between carriers and magnetic ions is dominated by the heavy hole contribution the decrease of the heavy-hole density of states results in the decrease of the Curie temperature.


We show that the Curie temperature may be dramatically changed by a small variation of the gate voltage in specially designed doped diluted semiconductor quantum wells. Taking into account the peculiarities of the complex valence band in a zinc-blende semiconductor,
we estimated that one can reach the Curie temperature variation by a factor of two with a variation of the applied voltage by about 5V. The predicted effect is caused by the step-like variation of the density of states in the valence band in the vicinity of a Lifshitz topological transition that can be controlled by the external bias. A strong sensitivity of the Curie temperature to the applied voltage can be used in semiconductor spintronic devices, in particular in ferromagnetic switches and spin transistors.

\begin{acknowledgments}
MMG is grateful to RSF project \# 17-12-01265 for partial support. A.V. and A.K. acknowledge partial support from the
HORIZON 2020 RISE project CoExAn (Grant No. 644076).
\end{acknowledgments}


\begin{thebibliography}{99}
\bibitem{note:J} For the homogeneous distribution of Mn ions the quantities $J_{hh}$, $J_{lh}$ are insensitive to the specific form of the envelope functions.
\end{thebibliography}
\end{document}